\documentclass[superscriptaddress,showpacs,amsmath,amssymb]{revtex4}

\usepackage{soul}

\usepackage[english]{babel}

\usepackage{braket}
\usepackage{tikz}
\usepackage{relsize}
\usepackage{framed}
\usepackage{bm}
\usepackage{graphicx,color}
\usepackage{hyperref}

\usepackage{amsfonts}
\usepackage{mathrsfs}

\newcommand{\s}{\mbox{\tiny S}}
\newcommand{\B}{\mbox{\tiny B}}

\newcommand{\M}{\mbox{\tiny M}}
\newcommand{\D}{\mbox{\tiny D}}
\newcommand{\tL}{\mbox{\tiny L}}
\newcommand{\tR}{\mbox{\tiny R}}
\newcommand{\ti}{\Tilde}

\newcommand{\nl}{\nonumber \\}
\newcommand{\nla}{\nl&\quad}

 \newcommand{\lgter}{\mbox{\tiny $\lessgtr$}}

\newcommand{\Sec}[1]{Sec.\,\ref{#1}}
\newcommand{\App}[1]{Appendix\,\ref{#1}}
\newcommand{\be}{\begin{equation}}
\newcommand{\ee}{\end{equation}}
\newcommand{\bea}{\begin{eqnarray}}
\newcommand{\eea}{\end{eqnarray}}
\newcommand{\bsube}{\begin{subequations}}
\newcommand{\esube}{\end{subequations}}
\newcommand{\Eq}[1]{Eq.\,(\ref{#1})}
\newcommand{\Eqs}[1]{Eqs.\,(\ref{#1})}
\newcommand{\Fig}[1]{Fig.\,\ref{#1}}

\newcommand{\dg}{\dagger}
\newcommand{\la}{\langle}
\newcommand{\ra}{\rangle}

\begin{document}

\title{ Distinguishing Majorana bound states from Andreev bound states  
through differential conductance and current noise spectrum}

 
\author{Huajin Zhao}
\affiliation{School of Physics, Hangzhou Normal University,
Hangzhou, Zhejiang 311121, China}

\author{Junrong Wang}
\affiliation{School of Physics, Hangzhou Normal University,
Hangzhou, Zhejiang 311121, China}

\author{Hong Mao} \email{mao@hznu.edu.cn}
\affiliation{School of Physics, Hangzhou Normal University,
 Hangzhou, Zhejiang 311121, China}

\author{Jinshuang Jin} \email{jsjin@hznu.edu.cn}
\affiliation{School of Physics, Hangzhou Normal University,
Hangzhou, Zhejiang 311121, China}

\date{\today}

\begin{abstract}
We investigate the quantum transport through a quantum dot coupled with a superconducting
(SC) nanowire. By elaborating the differential conductance and current noise spectrum,
we focus on the distinct characteristics
of the topological Majorana bound states (MBSs) and 
 trivial Andereev bound states (ABSs) hosted in SC wire.
For MBSs with a topological quality factor $q=1$,
we observe the degenerate features
manifested as the zero-bias peak (ZBP) 
in differential conductance and the Rabi dips degeneracy (RDD) in noise spectrum.
In contrast, for ABSs with $q<1$,
the splitting of these degenerate features  
depends on the linewidth, arising from realistic measurement conditions.
 Furthermore,
we identify the critical quality factors 
$q_{\rm c}$ and $q_{\s}$ associated with the emergences of ZBP and
  RDD, respectively.
The value of $q_{\rm c}$ is temperature-dependent, and we 
establish a suitable temperature window to ensure the visibility of single ZBP in the experiments. 
Whereas, $q_{\rm c}$ depends on the coupling strength rather than the temperature.
Typical values for these quality factors are
approximately $q_{\rm c}\approx 0.93$ and $q_{\s}\approx 0.99$.
Our results suggest that 
the degenerate Rabi spectrum signal  
 could serve as a hallmark for the presence of MBSs, 
 which goes beyond the scope of differential
conductance.
 
\end{abstract}



\maketitle

\section{Introduction}

Motivated by the promising applications in topological quantum computations
\cite{Kit01131,Kit032,Nay081083,Lei11210502},  
Majorana bound states (MBSs) have attracted extensive attention 
 in both theoretical \cite{Lut10077001,Ore10177002,
Fle10180516,Sar16035143,Moo18165302,Law09237001, 
Dan20036801,Men20036802}
and experimental studies
\cite{Mou121003,Den126414,Das12887,  
Fin13126406,Alb161038}.
Various realizations of the physical Majorana platforms have been proposed 
\cite{Lut10077001,Ore10177002,
Fle10180516,Sar16035143,Moo18165302,Law09237001, 
Dan20036801,Men20036802},
with the spin-orbit coupling
semiconductor nanowires proximity coupled to a superconductor emerging as a particularly popular scheme.
In this platform, %
zero-energy MBSs are expected to reside at the ends of the nanowire.
\cite{Lut10077001,Ore10177002}.
Experiments are designed to validate the existence of MBSs,
but the evidence remains controversial.
Tunneling spectroscopy in transport measurements has been 
 explored  
in both single-lead (two-terminal) 
\cite{Fle10180516,Sar16035143,Moo18165302,Law09237001},  
and two-lead (three-terminal) configurations \cite{Dan20036801,Men20036802}.
However, the observations of the zero-bias conductance peak and its quantized value $2e^2/h$
are insufficient for conclusively identifying the presence of MBSs
\cite{Liu17075161,Awo19117001,Avi19133,San1621427,Cay15024514}.
Therefore, a more powerful approach for inferring Majorana signatures
involves detecting nonlocality, such as through quantum correlation \cite{Bol07237002,Nil08120403,Qin22017402,Cao12115311,
Zoc13036802,Lu14195404,Lu16245418,Fen21123032},  
 or
 interferometry \cite{Sau15020511,Hel18161401}.  
Proposed hybrid setups, where MBSs are coupled with two quantum dots,
 present intriguing possibilities \cite{Zoc13036802,
Lu14195404,Lu16245418,Fen21123032}.
Yet, realizing these schemes 
remain great challenges to current experiments.  
So that an alternative scheme which is more realizable is still in demand.

Recently, it has been suggested that 
the nonlocal behavior of a topological system may
be assessed through a local measurement      
with the aid of a quantum dot (QD) at the wires's end \cite{Cla17201109,Pra17085418},
as illustrated in \Fig{fig1}.
Specifically, the Majorana non-locality is characterized by a``topological 
quality factor $q$'', defined as \cite{Cla17201109} $q=1-|\lambda_2|/|\lambda_1|$,
where $\lambda_1$ and $\lambda_2$
represent the couplings 
between the two Majorana modes and the QD.
The topological quality factor $q$ can be derived from differential conductance measurements via
the QD,
as experimentally demonstrated in Ref.\,\cite{Den18085125}.
 The observed zero-bias conductance peak, aligning with a high-quality factor approaching 1
($q\rightarrow1$), is indicative of purely topological MBSs hosted
in the SC wire. In contrast, the peak splitting associated with low-quality factor ($q<1$)
 suggests the presence of trivial Andereev bound states (ABSs).
Notably,
  ABSs with partially separated Majorana states, termed quasi-MBSs (qMBSs)
 with $|\lambda_2|\ll|\lambda_1|$,
 can also exhibit high-quality factors,  
  mimicking the signatures of spatially separated topological MBSs
\cite{Liu18214502,Vui19061}.
This similarity complicates the disinction
between qMBSs and topological MBSs
 through the proposed local measurements \cite{Cla17201109,Pra17085418}.
Consequently, two fundamental questions arise:   
What is the critical quality factor of qMBSs that mimic the characteristics of MBSs,
and what the physical mechanisms determine this critical quality factor?
Addressing these issues will be instrumental for advancing current experimental demonstrations.


In this paper, we will thoroughly study the quantum transport through a QD 
 coupled with a superconducting (SC) nanowire,
 examining both the differential conductance and the current noise spectrum.
 %
 As is well-known, shot noise stemming from nonequilibrium quantized charge
current fluctuations offers profound insights
into  nonequilibrium quantum transport beyond the average current
or the differential conductance
\cite{Her927061,Bla001,Imr02,Naz03}.
In particular, the full frequency current noise spectrum can reveal
the information about the intrinsic coherent
dynamics of the system \cite{Ent07193308,Li05066803,Bar06017405,Wab09016802,Jin13025044,Jin11053704,
Rot09075307,Yan14115411,Shi16095002,Jin20235144,Xu22064130}.
Recently, research on the Majorana-related noise in similar transport setups 
 has garnered attention \cite{Liu15081405,Smi19165427,Smi22205430,Fen22035148,Cao23121407}.
For instance, Smirnov demonstrated the Majorana universality of the differential
shot noise at low bias voltages \cite{Smi22205430} and exhibited
the universal Majorana finite-frequency features
   at strong coupling regimes \cite{Smi19165427}.
 Cao {\it etal.,} proposed identifying ABS-induced quantized conductance plateaus
by measuring the associated differential shot noise as a function of bias voltage \cite{Cao23121407}.
%
Here, based on both differential conductance and current noise spectrum, 
we first elucidate the distinct nonequilibrium transport characteristics 
between topological MBSs and trivial ABSs hosted in the SC wire.
Subsequently, 
we explore how qMBSs 
 with high-quality factors, exceeding the critical values, 
 mimic the signatures of spatially separated topological
MBSs. Importantly, 
we reveal the underlying mechanism governing the critical quality factors
 for the emergence of the zero-bais peak in differential conductance and the
Rabi signal degeneracy in noise spectrum,
respectively.
Finally, we provide quantitative expressions for the
critical quality factors, which could be instrumental for current experimental demonstrations.

The remainder of this paper is organized as follows.
In \Sec{thMet}, we introduce the hybrid transport model and 
the employed quantum master equation approach.
The related superoperators involved in current expression and the noise formula are
defined in \App{appsuper}.  
We present the results in \Sec{thres}.
First, we analyze the QD-wire Hamiltonian eigenspectrum in \Sec{thHam}.
Then we demonstrate the detail results for the differential conductance and the current noise spectrum
in \Sec{thdotI} and \Sec{thdotII}, respectively.
Finally, we give the summary in \Sec{thsum}.

\section{Methodology}
\label{thMet}

\subsection{Model description}

\begin{figure}
\includegraphics[width=0.5\columnwidth]{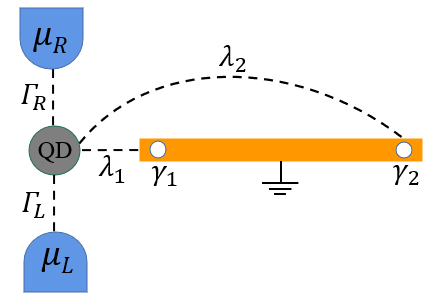}  
\caption{(Color online)
Schematic diagram for the transport through the QD-wire hybrid system.
The QD is contacted by the two electron reservoirs under the bias voltage
($V=\mu_{\tL}-\mu_{\tR}$) and laterally coupled to the two Majorana modes
($\hat\gamma_1$ and $\hat\gamma_2$).
They are hosted in
the superconducting wire
with the coupling coefficients $\lambda_1$ and $\lambda_2$, respectively.
  }
\label{fig1}
\end{figure}
We consider the electron transport through a quantum dot (QD) which is laterally coupled
to a superconducting (SC) nanowire as schematically shown in \Fig{fig1}.
The composed Hamiltonian consists of three parts,
$H_{\rm tot}=H_{\B}+H'+H_{\s}$.
The first part is the bath Hamiltonian of the two electron reservoirs
which is given by $H_{\B}=\sum_{\alpha}(\varepsilon_{\alpha k}+\mu_\alpha)
\hat c^\dg_{\alpha k}\hat c_{\alpha k}$, with
the applied bias voltage $V=\mu_{\tL}-\mu_{\tR}$.
The second part is the coupling Hamiltonian between the reservoirs and QD. It is
described by the standard tunneling form,
  \begin{align}\label{Hcoup}
  H'&=  \sum_{\alpha={\rm L,R}}\hat F^\dg_\alpha\hat d+  \hat d^\dg \hat F_\alpha,
 \end{align}
 where $\hat d^\dg$ ($\hat d$) is the electron creation (annihilation) operator
 in the QD with the assumption of a single fermionic mode, and
 $\hat F^\dg_\alpha\equiv\sum_{ k} t_{\alpha k}  \hat c^\dg_{\alpha k}$
with the tunneling coefficients $t_{\alpha k} $ and $\alpha={ L,R}$.
 Throughout this work, we adopt units of $e=\hbar=1$
for the electron charge and the Planck constant.
 Introduce also the sign symbol $\sigma=\pm$ and its opposite sign $\bar\sigma$,
for such that $\hat d^+\equiv \hat d^\dg$ and $\hat d^-\equiv \hat d$.

The primary interest is the QD-wire hybrid system in
 the third part of the composed Hamiltonian. 
 It can be simply described by the low energy Hamiltonian \cite{Cla17201109},
\begin{align}\label{HS0}
  H_{\s}\!=\!\varepsilon_{\D}\hat d^\dg \hat d +i \varepsilon_{\M} \hat\gamma_{1} \hat\gamma_{2}
  +\big(\lambda_{1} \hat d^\dg  \hat\gamma_{1}  +i \lambda_{2} \hat d^\dg \hat\gamma_{2} +{\rm H.c.}\big).
 \end{align}
 Here, the first term is the Hamiltonian of the QD 
 with the energy level $\varepsilon_{\D}$.
 The wire hosts a pair of Majorana modes $\hat\gamma_1$ and $\hat\gamma_2$
 with the coupling energy $\varepsilon_{\M}$.
 The last term in \Eq{HS0} describes the coupling between the dot and two Mojorana modes
 with the two different coupling coefficients $\lambda_1$ and $\lambda_2$. 
 Without loss of generality,
 we assume $\lambda_1$ and $\lambda_2$ to be positive real.
%

The quality factor is thus defined as $q=1-\frac{\lambda_2}{\lambda_1}$ \cite{Cla17201109}.
 If the wire is in the topological regime, $\lambda_2$ should be zero ($\lambda_2\rightarrow0$) and the quality factor
 approaches one ($q\rightarrow 1$).
 This means that there are two MBSs with the modes
  $\hat\gamma_1$ and $\hat\gamma_2$ located at the two
 ends of the wire.  
 Otherwise ($\lambda_2\neq0$ and $q < 1$), the wire is in the nontopological regime 
 where the mode $\hat\gamma_2$ is close to mode $\hat\gamma_1$.
 In this case, the wire bears trivial ABSs
  including quasi-MBSs (qMBSs) as  
 $\hat\gamma_1$ and $\hat\gamma_2$ are partially separated with $\lambda_2\ll\lambda_1$
  \cite{Chi18035312,Moo18155314,Liu18214502,Vui19061}.
Furthermore, for the case of zero quality factor ($q=0$ with $\lambda_2=\lambda_1$), 
the two modes of $\hat\gamma_1$ and $\hat\gamma_2$ locate the same position and form a regular fermion.
This leads to the disappearance of Andereev process in the dot-wire Hamiltonian of \Eq{HS0},
as formulated explicitly 
in \Eq{HSapp}.
In this work, our study will focus on the different 
 characteristics of the differential conductance and current fluctuation spectrum 
 between MBSs and ABSs hosted in SC wire. 
In particular, we will explore the dependence of these transport characteristics 
on the quality factor.

 \subsection{Quantum master equation approach}

For the consideration of the weak QD-reservoir coupling[c.f.\Eq{Hcoup}],
we apply the second-order quantum master equation (QME) with memory effect. It
is described by the time-nonlocal QME \cite{Jin11053704,Yan05187,Mak01357,Jin16083038},
\be\label{QME}
\dot\rho(t)\! =\! -i{\cal L}_{\s}\rho(t)\!-\!\sum_{\alpha \sigma}
  \!\int_{t_0}^t\!\!{\mathrm d}\tau \big[\hat d^{\bar\sigma},
  {\cal C}^{(\sigma)}_{\alpha }({\cal L}_{\s},t-\tau) \rho(\tau) \big],
 \ee
where ${\cal L}_{\s} \bullet=[H_{\s},\bullet]$ and
\be\label{calCt}
 {\cal C}^{(\sigma)}_{\alpha }({\cal L}_{\s} ,t-\tau) \bullet
 \equiv    e^{-i{\cal L}_{\s} (t-\tau)} \big[c^{(\sigma)}_{\alpha }(t-\tau)\hat d^\sigma \bullet
 - c^{(\bar\sigma)\ast}_{\alpha }(t-\tau)\bullet \hat d^{\sigma} \big],
 \ee
with $c^{(\sigma)}_{\alpha }(t-\tau)=\la \hat F^\sigma_{\alpha }(t)\hat F^{\bar\sigma}_{\alpha }(\tau)\ra$
being the bath correlation function.
In \Eq{QME}, the first term describes the intrinsic coherent dynamics
of the dot-wire hybrid system. The second term  
depicts the non-Markovian dissipative effect of the coupled electron reservoirs.

The current through the system via the definition of $ I_\alpha(t)=\la\hat I_\alpha(t)\ra$, with 
$\hat I_\alpha(t)=-{\rm d}\hat N_\alpha (t)/{\rm dt}$ and
$\hat N_\alpha\equiv \sum_k \hat c^\dg_{\alpha k}\hat c_{\alpha k}$, is given by \cite{Jin11053704,Shi16095002,Xu22064130}
\be\label{curr-exp}
  I_{\alpha}(t)
=-\!\sum_{\sigma u} \!\!\int_{t_0}^t\!\! {\mathrm d}\tau\, {\rm tr}_{\rm s}
   [\ti d^{\sigma} {\cal C}^{(\bar\sigma)}_{\alpha }({\cal L}_{\s} ,t-\tau) \rho(\tau) ],
\ee
where $\ti d^{\sigma}=\sigma \hat d^{\sigma}$.
  The stationary current can be further expressed with
  $\bar I_{\alpha}=\big[{\cal J}^{>}_{\alpha}(0)\bar{\rho}\big]
  ={\rm tr}_{\s}\big[{\cal J}^{<}_{\alpha}(0)\bar{\rho}\big]$.
  Here, the superoperators ${\cal J}^{\lgter}_{\alpha}$ is defined in \Eq{caljomega}
  and
  $\bar \rho$ is the steady state of the dot-wire hybrid system which is 
  the solution of 
  $\dot\rho(t)=0$ in \Eq{QME}.

 The current noise spectrum, 
 $  S_{\alpha\alpha'}(\omega)=\int {\rm d}t\,
     e^{i\omega t}  \la \delta{\hat I}_\alpha(t)
     \delta{\hat I}_{\alpha'}(0)\ra$
 with $\delta{\hat I}_\alpha(t)\equiv{\hat I}_\alpha(t)-\bar I_{\alpha}$,
can be calculated via \cite{Xu22064130}
\begin{align}\label{Sw}
 S_{\alpha\alpha'}(\omega)
&=2\delta_{\alpha'\alpha}{\rm Re}\! \sum_{\sigma  }
   {\rm tr}_{\rm s}\big[\hat d^{\bar\sigma}
   c^{\sigma}_{\alpha  }(\omega-{\cal L}_{\s})
   (\hat d^{\sigma} {\bar\rho)}\big]
\nl&\quad
 +{\rm tr}_{\rm s}\Big\{
  {\cal J}^{<}_{\alpha'}(-\omega){\cal G}(-\omega)\big[{\cal J}^{<}_{\alpha}(0)  +
  {\cal W}^{<}_{\alpha}(-\omega) \big]\bar\rho
\nl&\qquad
 +{\cal J}^{>}_{\alpha}(\omega)
 {\cal G}(\omega)\big[{\cal J}^{>}_{\alpha'}(0)
   + {\cal W}^{>}_{\alpha'}(\omega)\big]\bar\rho
\Big\},
 \end{align}
 with the related superoperators being defined  in the \App{appsuper}.
This describes the
  energy absorption ($\omega>0$) and emission ($\omega<0$) processes 
  \cite{Eng04136602,Rot09075307,Jin15234108}.

%

\section{Results}
\label{thres}
\subsection{Eigenspectrum with parity analysis}
\label{thHam}

As well-known, the Majorana modes are related to the regular fermion through 
the transformation of $\hat\gamma_{1} =\hat f +\hat f^\dg $ and $\hat\gamma_2 = -i \big(\hat f - \hat f^\dg \big)$.
The system Hamiltonian of \Eq{HS0} can be rewritten as   
 \begin{align}\label{HSapp}
  H_{\s}&=\varepsilon_{\D} \hat d^\dg \hat d +\varepsilon_{\M} (\hat f^\dg \hat f-\frac{1}{2})  
 \!+\! \big(\lambda_1\!+\!\lambda_2\big)  \hat f^\dg \hat d\!+\!{\rm H.c.}\big]
 \!+\! \big[\big(\lambda_1\!-\!\lambda_2\big) \hat f\hat d\!+\!{\rm H.c.}\big].
 \end{align}
 The last two terms describe
 the normal tunneling and Andreev reflection processes, respectively.
 In the Fock states basis $\{|n_{\D}n_{\M}\ra=|01\ra,|10\ra,|00\ra,|11\ra\}$,
 associated with 
 $ \hat n_{\D}= \hat d^\dg  \hat d $
and $ \hat n_{\M}= \hat f^\dg \hat f $,
\Eq{HSapp} is block diagonal, 
 \begin{equation}\label{Hsm}
H_{\s}={
  \left( \begin{array}{cc}
           H_{-} & 0 \\
           0 & H_{+}
         \end{array}
         \right)},
\end{equation}
in the odd-parity ($p=-1$) 
 and even-parity ($p=+1$) subspaces,
with
\bsube
\label{Hfock}
\begin{equation}
H_{-}={
\left( \begin{array}{cc}
         \varepsilon_{\M} & \lambda_{1}+\lambda_{2}\\
        \lambda_{1}+\lambda_{2} & \varepsilon_{\D}
       \end{array}
       \right)},
\end{equation}
\begin{equation}
H_{+}={
\left( \begin{array}{cc}
         0 & \lambda_{1}-\lambda_{2}\\
        \lambda_{1}-\lambda_{2} & \varepsilon_{\D}+\varepsilon_{\M}
       \end{array}
       \right)}.
\end{equation}
\esube
 Apparently, there are intrinsic coherent Rabi oscillations in the odd and even parity subspaces
 which are closely related to 
 the Majorana qubit readout \cite{Gha16155417,Mun20033254,Ste20033255,Khi21127}.
The corresponding parity-dependent Rabi-frequency reads
\be\label{deltap0}
\Delta_{\rm p}=\sqrt{(\varepsilon_{\D}+{\rm p}\varepsilon_{\M})^2
  +4 (\lambda_1-{\rm p}\lambda_2)^2}.
\ee
As known, the noise spectrum, \Eq{Sw}, exhibits 
some obvious characteristics at Rabi frequency $\Delta_{\rm p}$.
For instance, it has been demonstrated that a dip feature occurs at Rabi frequency
in the noise spectrum
in both serial and parallel coherent
coupled double dots \cite{Luo07085325,Don08033532,Mar11125426,Shi16095002}.
 Hence, we could expect that the parity-dependent Rabi 
 signals at $\omega\simeq \pm\Delta_{\rm p}$, via \Eq{deltap0} with ${\rm p}=\pm$,
 should emerge in the current noise spectrum.

 On the other hand, the conductance peaks
 appear whenever the Fermi surface of the electrode is on resonance with the allowed transitions
 between the ground state and
 an excited state with the opposite parity \cite{Cla17201109}. 
The resulting peak positions are expressed 
in descending order as
\be\label{peaks}
\begin{split}
 E_{1} &=\frac{1}{2}(\Delta_{+}+\Delta_{-}),~~
    E_{2}=\frac{1}{2}(-\Delta_{+}+\Delta_{-}),
\\
 E_{3}  &=\frac{1}{2}(\Delta_{+}-\Delta_{-}),~~
 E_{4}  =-\frac{1}{2}(\Delta_{+}+\Delta_{-}).
\end{split}
\ee
From \Eqs{deltap0} and (\ref{peaks}), it is easy to find that the zero-bias 
peak (ZBP) occurs at $\mu_{\rm L}=V/2=E_2=E_3=0$, satisfying the condition of
 $\Delta_+=\Delta_-$ that can be recast in terms of
\begin{align}
\label{zbpc}
\varepsilon_{\D}\varepsilon_{\M}=4\lambda_1\lambda_2.  
 \end{align}
 For the wire in the topological regime ($q=1$ with $\lambda_2=0$),
we can have either $\varepsilon_{\D}=0$ or $\varepsilon_{\M}=0$.
In each situation, 
 the ZBP arises in the differential conductance,
 while the degenerate Rabi feature occurs at $\omega=\Delta_+=\Delta_-$ in the noise spectrum.
However, 
the ZBP of conductance 
splits into a pair for the trivial ABSs in
the nontopological regime ($q<1$ with $\lambda_2\neq0$); see \Fig{fig2}.
The similar features occur for 
the degenerate Rabi signal in the noise spectrum; see \Fig{fig4} and remarked therein.

The above analysis 
would imply that either the ZBP in the conductance or 
 the degenerate Rabi signal in the noise spectrum  
could be taken as a hallmark for the existence of MBSs in SC wire. 
However, for the realistic measurement setup as shown in 
 \Fig{fig1}, there are line broadenings and whether the degenerate features split or not depend on the linewidth. 

 Note that  
the ZBP occurs when $\Delta_+=\Delta_-$ in which \Eq{zbpc} holds.
To elaborate the distinct characteristics between the topological MBSs and trivial ABSs,
we focus on the case of $\varepsilon_{\D}\varepsilon_{\M}=0$. 
  We identify the critical quality factor 
$q_{\rm c}$ for the ZBP in the conductance and
also that
 $q_{\s}$ for the degenerate Rabi feature in the noise spectrum.
The typical values are $q_{\rm c}\approx 0.93$ and $q_{\s}\approx 0.99$.
In other words,
the degenerate Rabi spectrum signal  
 would be the hallmark of MBSs, which goes beyond the scope of differential conductance.

\subsection{The differential conductance}
\label{thdotI}

   \begin{figure}
\centering
\includegraphics[width=0.60\textwidth,clip=true]{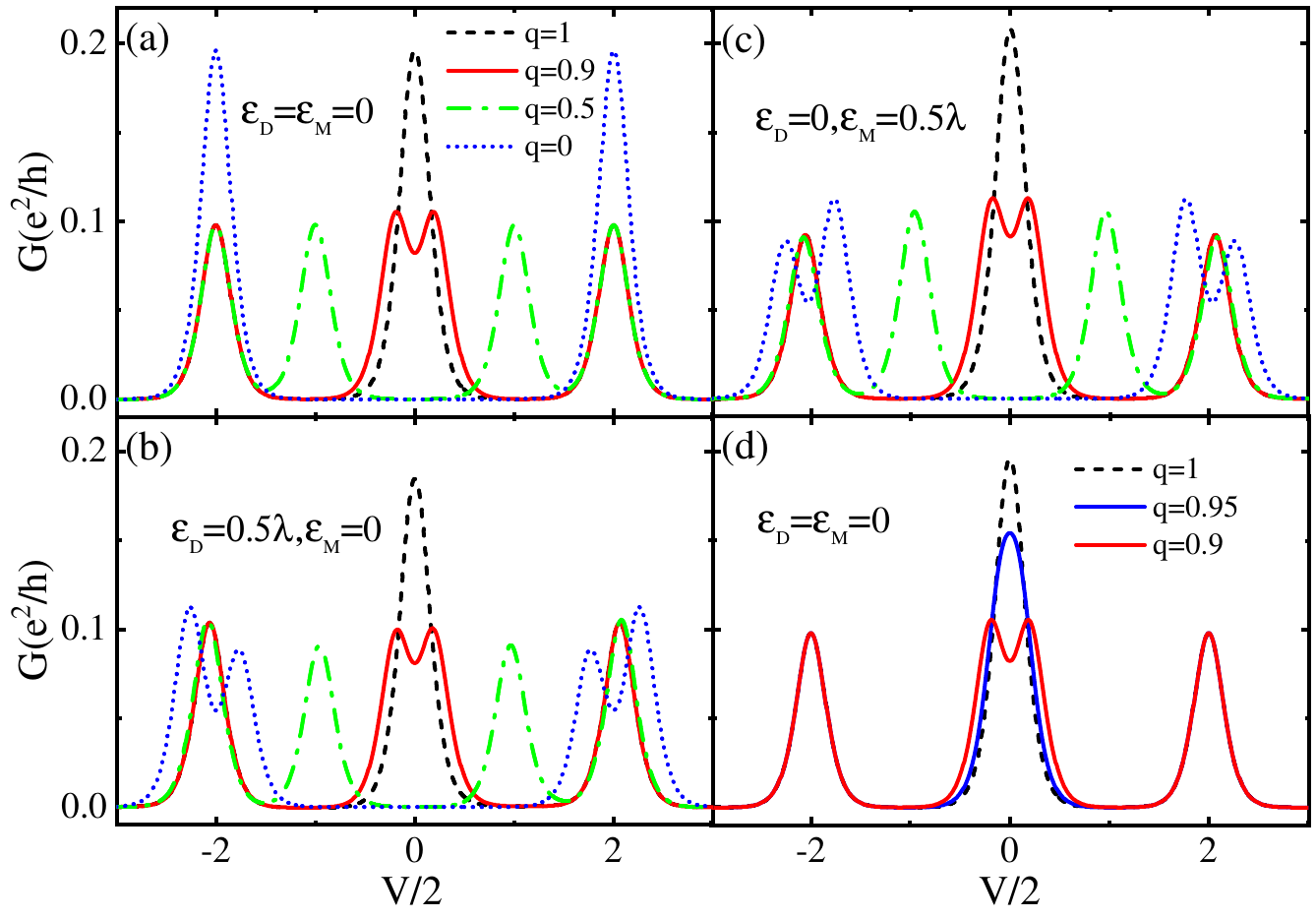}
 \caption{ (Color online) The differential conductance $G$ as a function of bias voltage $V$
 with different quality factor $q$. 
The other parameters are (in unit of $\lambda$): $k_{\B}T=0.1$ and $\Gamma=0.1$.
  \label{fig2}}
\end{figure}
 
We first demonstrate the characteristics of differential
conductance, especially the value of $q_{\rm c}$, in the transport measurement
setup of \Fig{fig1}.  
In our calculations, we consider the symmetrical-lead situation,
where $\mu_{\tL}=-\mu_{\tR}=V/2$ and $\Gamma_{\tL}=\Gamma_{\tR}= \Gamma/2$.
Consequently, we have
$G=G_{\rm L}=-G_{\rm R}=\frac{{\rm d}{\bar I}_{\rm L}}{{\rm d}V}$. 
  Furthermore, 
we set $\lambda_1=\lambda=1$ as a unit of the energy level 
and then $\lambda_2=(1-q)\lambda$. 
We also consider the wide bandwidth limit $W\gg\Gamma$ for the electron reservoirs.
The parameters of the coupling strength $\Gamma$ and the electrode temperature $k_BT$
 are adopted in the weak coupling regime, where $\Gamma\lesssim k_BT$.

Figure \ref{fig2} illustrates the
differential conductance as a function of the bias voltage,
 in which the conductance peaks occur at $|\mu_{\alpha}|=V/2=E_{i}$, with $i=1,2,3,4$.
As demonstrated,
the conductance displays three peaks
for $q=1$ (MBSs). The height of the ZBP is twice that of the other two peaks due to
the overlap of two peaks ($E_2=E_3=0$) at zero-bias voltage.
For the wire hosting ABSs with $q<1$,
the ZBP splits ($E_2\neq E_3\neq0$), leading to four peaks.
For the regular local fermion with $q=0$,  
 two degenerate peaks appear at $E_{1}=E_{2}$ and $E_{3}=E_{4}$
 when $\varepsilon_{\M}=\varepsilon_{\D}=0$.
 On the contrary, 
 for either $\varepsilon_{\D}\neq0$  or $\varepsilon_{\M}\neq0$,
 the degenerate peaks split, resulting in four peaks as displayed in 
 \Fig{fig2}(b) and \Fig{fig2}(c).
The heights of these four peaks are different,
due to the nonsymmetric tunneling channels with $\varepsilon_{\M}\neq\varepsilon_{\D}$.

So far, we have understood that the key to distinguish ABSs from MBSs based on the conductance
characteristics lies in observing whether the ZBP splits. Looking at \Fig{fig2}(a)-(c),
 the ZBP 
  displays the certain broadening that completely originates from 
the dissipative effect of the contacted electron reservoirs on the QD-wire system.
This broadening, caused by the measurement, limits the distinction efficiency between MBSs and ABSs. 
Thus, the ZBP in the conductance cannot be the hallmark 
 for MBSs without considering any other physical mechanisms. 
For example, we plot the differential conductance for a high-quality factor $q=0.95$ (qMBSs) in \Fig{fig2}(d).
It indicates that the qMBSs indeed
  mimic the ZBP signature of topological MBSs, as demonstrated in previous work
\cite{Liu18214502,Vui19061}.

 \begin{figure}
\centering
\includegraphics[width=0.46\textwidth,clip=true]{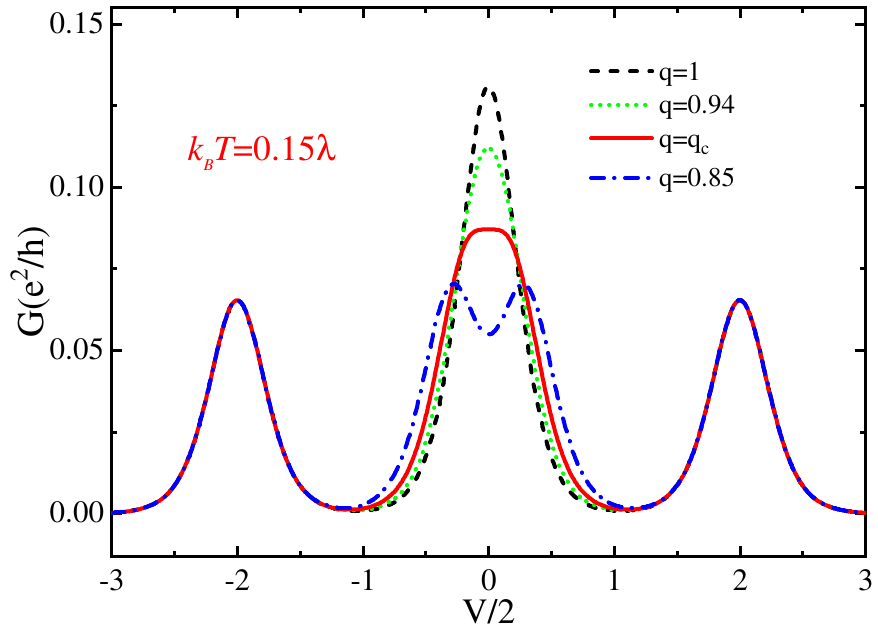}
 \caption{ (Color online) The differential conductance $G$ as a function of bias voltage $V$
 with different quality factor ($q$) for the demonstration
of the critical quality factor $q_{\rm c}$.
For the parameters used here, $q_{\rm c}\approx 0.934$.
The other parameters are (in unit of $\lambda$):
 $k_{\B}T=0.15$, $\varepsilon_{\D}=\varepsilon_{\M}=0$ and $\Gamma=0.1$.
  \label{fig3}}
\end{figure}

 Now we will explore the critical quality factor 
  $q_{\rm c}$ of qMBSs that mimics the ZBP characteristic of MBSs.
To that end, we provide an
 analytical expression using $\varepsilon_{\D}=\varepsilon_{\M}=0$
as an example. The resulting differential conductance is given by
\begin{align}\label{G00}
  G(V)&=\frac{\Gamma \beta}{2^7}
  \Big({\rm sech}^2[\beta(V/4-\lambda_1)]+{\rm sech}^2[(V/4+\lambda_1)]
 \nl& 
  \!+\!{\rm sech}^2[\beta(V/4-\lambda_2)]\!+\!{\rm sech}^2[\beta(V/4+\lambda_2)]\Big),
 \end{align}
where we have set $\lambda_1=\lambda$ and $\lambda_2=(1-q)\lambda$.
Obviously, the conductance in \Eq{G00} indicates
that the peaks arise at the bias voltage of $V/2=E_1=2\lambda_1=-E_4$ and
$V/2=E_2=2\lambda_2=-E_3$, which are consistent with those given by \Eq{peaks}.
Moreover, the height of the peaks is affected by both the coupling strength and temperature, whereas
the width of the peaks is solely determined by the temperature. 
From \Eq{G00}, it is easy to obtain the half-width  
 of the peaks as 
 \be\label{gammad}
 \Gamma_{\rm d}=2\ln{(1+\sqrt{2})}\,k_{\B}T\approx 1.763 \,k_{\B}T.
 \ee 
 As mentioned above, the ZBP at $E_2=E_3=0$ for $q=1$ should symmetrically split into two peaks at finite values 
  of $E_2$ and $E_3=-E_2$ for $q<1$ (ABSs), as illustrated in \Fig{fig2} with the example of $q=0.9$.
  This implies the existence of a critical quality factor $q_{\rm c}$,
   which corresponds to an inflection point in the differential
  conductance at $V=0$.
  Based on \Eq{G00}, one then easily gets the critical quality factor as 
  \be\label{qc}
  q_{\rm c}=1-\ln{(2+\sqrt{3})}\frac{ k_{\B}T}{2\lambda}
  \approx1-0.66 \frac{ k_{\B}T}{\lambda}.
  \ee  
  This suggests that
  we can only distinguish the conductance features between  
  $q> q_{\rm c}$ and $q< q_{\rm c}$. The former displays ZBP and 
  the latter exhibits ZBP splitting,
  as numerically illustrated in \Fig{fig3}.
  Such a result could explain 
  why
  the partially separated of qMBSs
  with high-quality factor ($q> q_{\rm c}$ with $\lambda_2\ll\lambda_1$)  
 mimic the characteristics of
MBSs (q = 1) \cite{Liu18214502,Vui19061}. Therefore, we cannot distinguish qMBSs from
 spatially separated topological MBSs based on the characteristics 
of differential conductance.
  %

  %
Note that the obtained $q_{\rm c}$ in \Eq{qc} is actually applicable to 
the cases where $\varepsilon_{\M}=0$ or $\varepsilon_{\D}=0$,
and even very small values
of $\varepsilon_{\M},\varepsilon_{\D}\ll\lambda$.
However, considering the actual situation in experiments,
 it is preferable to tune the energy level of the dot as $\varepsilon_{\D}\rightarrow0$, 
 since the coupling energy ($\varepsilon_{\M}$) of Majorana modes is determined by the
length of the wire.

Furthermore, we emphasize that the present work focuses on the weak system-reservoir coupling regime,
 where $\Gamma\lesssim k_{\rm B}T$. 
  However, the ZBP will be smoothed by increasing the temperature, with its
  half-width broadening given by \Eq{gammad}.
Considering the measurement visibility of a single ZBP, the distance between
the ZBP and its nearby symmetric peaks should be greater than the broadening width,
i.e., $|E_1|=|E_4|>2\Gamma_{\rm d}$. This results in $k_{\rm B}T<\frac{\lambda}{2\ln{(1+\sqrt{2})}}$. 
Therefore, we identify the valid temperature window as
\be\label{temp}
\Gamma\lesssim k_{\B}T 
 <\frac{\lambda}{2\ln{(1+\sqrt{2})}},
\ee
 for the experimental observation of the ZBP.

\subsection{The noise spectrum }
\label{thdotII}
   \begin{figure}
\centering
\includegraphics[width=0.8\textwidth,clip=true]{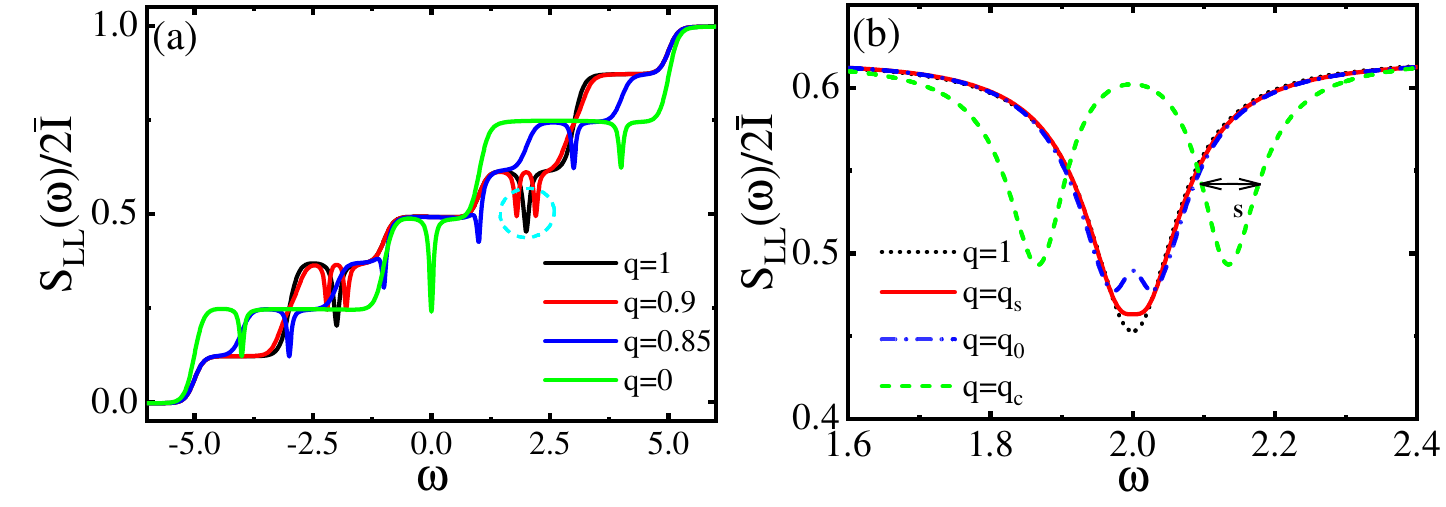}
 \caption{ (Color online) The auto-correlation current noise spectrum of the left-lead,
 $S_{\rm LL}(\omega)$, with different quality factors, for (a) the whole frequency-regime and (b)
 the regime around the Rabi frequencies. The value of $q_{\rm c}\approx 0.934$ is the same as in 
 \Fig{fig2}. Here $q_{0}\approx0.9885$ and $q_{\s}\approx0.994$.
 The other parameters are (in unit of $\lambda$)
 $\varepsilon_{\M}=\varepsilon_{\D}=0$, $\Gamma=0.1$, $k_{\B}T=0.1$, and $V=6$.
  \label{fig4}}
\end{figure}

We are now in the position to demonstrate the characteristics in the current noise spectrum,
with a particular focus on exploring the value of $q_{\s}$.
As known, the noise spectrum
contains the information beyond the conductance \cite{Her927061,Bla001,Imr02,Naz03}.
The differential conductance reflects the energy transition (or structure)
information of the QD-wire system as illustrated above.
In contrast, the current noise spectrum is related to
 not only the energy transitions but also the intrinsic coherent
dynamics of the system \cite{Ent07193308,Li05066803,Bar06017405,Wab09016802,Jin13025044,Jin11053704,
Rot09075307,Yan14115411,Shi16095002,Jin20235144,Xu22064130}.
For the former,  
it is manifested as the non-Markovian quasi-steps at
resonance frequencies,
$\pm|\mu_{\alpha}-E_i|$, as depicted in \Fig{fig4} (a).
This feature arises from
the non-Markovian dynamics of the electrons in $\alpha$-electrode
tunneling into and out of the system, accompanied by energy
absorption ($\omega>0$) and emission ($\omega<0$), respectively.
The corresponding differential characteristics ($\frac{dS_{\rm LL}(\omega)}{d\omega}$),
display peaks at $\pm|\mu_{\alpha}-E_i|$, which resemble
  those observed in the differential conductance \cite{Mao21014104}.
%

Of particular interest is the characteristic of the intrinsic 
coherent dynamics in the noise spectrum.
It exhibits 
  the coherent Rabi dip signal at $\omega\approx\pm\Delta_{\rm p}$,
  as illustrated in \Fig{fig4} (a). 
This attribute is derived from the propagator ${\cal G}(\omega)$ [c.f.
  \Eq{Gomega}] in the second and third terms of \Eq{Sw}.
Akin to the conductance,
the noise spectrum displays the similar feature 
for either $\varepsilon_{\M}=0$ or $\varepsilon_{\D}=0$. 
 For simplicity and without loss of generality, we use  
 $\varepsilon_{\M}=\varepsilon_{\D}=0$ as an illustrative example in
 subsequent discussions.
 %

As shown in \Fig{fig4} (a), $S_{\rm LL}(\omega)$ displays a deep Rabi dip, 
which we refer to as Rabi dips degeneracy (RDD),
at $\Delta_+=\Delta_-$ for $q=1$ (MBSs).
The RDD is separated into two dips at $\Delta_+$ and $\Delta_- 
\neq\Delta_+$, respectively, for $0<q<1$ (ABSs).
When $q=0$ (i.e., $\lambda_1=\lambda_2$), the Rabi signal at $\Delta_{+}$ vanishes
for the disappearance of the Andreev reflection processes; see \Eq{HSapp}.
The resulting noise spectrum displays only one dip 
at $\Delta_-=2\sqrt{2}\lambda$
in the high frequency regime.

Therefore, the crux to distinguish ABSs from MBSs in the noise spectrum is to observe whether the RDD is split,
as indicated by the dashed circle in \Fig{fig4} (a).
Analogous to the ZBP feature in differential conductance, the Rabi dip 
in the noise spectrum exhibits a certain broadening
that may hinder the efficiency of distinguishing between MBSs and ABSs. 
Given the complexity of the analytical results, 
our discussions are primarily
based on the numerical calculations. 
 Our findings indicate that, 
unlike the conductance, the broadening ($\Gamma_{\s}$) of the Rabi dip in the 
noise spectrum is predominantly determined by
the coupling strength ($\Gamma$) rather than the temperature ($k_{\B}T$).
Specifically, we observe that $\Gamma_{\s}\approx 4\Gamma/5$, as denoted by the 
double arrow in \Fig{fig4} (b).
The critical quality factor $q_{\s}$ is consequently 
obtained as 
\be\label{qs}
q_{\s}\approx1-\frac{5\Gamma_{\s}}{72\lambda}\approx1-\frac{\Gamma}{18\lambda}.
\ee
It is higher than that of the Lorentz counterpart:
\be
q_{0}=1-  \frac{\Gamma_{\s}}{4\sqrt{3}\lambda}\approx1-\frac{\Gamma}{5\sqrt{3}\lambda}.
\ee
Figure \ref{fig4} (b)
illustrates that for $q> q_{\s}$, RDD is present, while for $q< q_{\s}$, RDD is split.
Intriguingly, in the regime of weak coupling and low temperature determined by \Eq{temp},
the critical quality factor $q_{\s}$ is typically higher than $0.99$. 
For the parameters used in \Fig{fig4} with $\Gamma=0.1\lambda$, we calculate $q_{\s}\approx0.994$.  
By further reducing the coupling strength $\Gamma$, $q_{\s}$ approaches $1$. 
In this sense, the RDD signal in the noise spectrum emerges as a potential hallmark
 for the existence of MBSs in SC wire. This suggests that by analyzing the noise spectrum, 
 one may be able to identify the presence of MBSs with high accuracy.

\section{Summary}
\label{thsum}

In summary, we have thoroughly studied the 
differential conductance and the noise spectrum
of the transport current through the quantum dot coupled 
with a SC wire. 
The study is based on the quantum master equation approach for 
weak system-reservoir coupling regime ($\Gamma\lesssim k_{\B}T$).
We analyzed the distinct nonequilibrium transport characteristics 
between the topological MBSs and trival ABSs
hosted in the SC wire. 
For the SC nanowire hosting topological MBSs with the topological quality factor $q=1$, 
we demonstrated the degenerate characteristics: the zero-bias peak (ZBP) in differential conductance and
the Rabi dips degeneracy (RDD) in noise spectrum.
In contrast, for the SC nanowire bearing ABSs with $q<1$, 
 whether the degenerate features split or
not depend on the linewidth, due to the line
broadenings inherent in realistic measurement setups.
Furthermore,
we identified the critical quality factors
$q_{\rm c}$ and $q_{\s}$ for the emergences of the ZBP in differential conductance and
 the RDD in current noise spectrum, respectively.

For the differential conductance, we derived an analytical expression
for $q_{\rm c}$. It enable us to distinguish the conductance features between  
  $q> q_{\rm c}$ and $q< q_{\rm c}$, exhibiting a ZBP and 
 ZBP splitting. 
The resulted $q_{\rm c}$ depends on
the temperature of the electron reservoirs.
In addition, 
  we identified a valid
temperature window for experimental  
measurements to ensure the visibility of a single ZBP

For the current noise spectrum, we numerically validated the critical quality factor 
$q_{\s}$ for RDD. 
Notably, $q_{\s}$ is determined by the system-reservoir coupling strength
rather than the temperature, and surprisingly approaches one ($q_{\s}\rightarrow 1$)
in the considered weak system-reservoir coupling regime.
Thus, beyond the scope of differential conductance,
the RDD signal in the noise spectrum 
would serve as a hallmark for the existence of MBSs in SC wires.
 We anticipate that our findings will provide valuable insights for ongoing experimental demonstrations.

\acknowledgments
We acknowledge helpful discussions Prof. YiJing Yan.
  The support from the Natural Science Foundation of China
(Grant No. 12175052) is acknowledged.

\appendix
\label{thapp}

\section{Superopertors in the noise spectrum formula}
\label{appsuper}
The detail derivation for the noise spectrum formula \Eq{Sw}
can refer to Ref.\,\onlinecite{Xu22064130}. 
The involved superoperators in \Eq{Sw} read,
\bsube
 \label{caljwomega}
 \begin{align}\label{caljomega}
{\cal J}^{>}_{\alpha}(\omega)\hat O&\equiv\!-\!\sum_{\sigma  }\ti{d}^{\bar\sigma} 
\big[{\cal C}^{(\sigma)}_{\alpha  }(\omega-{\cal L}_{\s})\hat O\big],
\\
{\cal J}^{<}_{\alpha}(\omega)\hat O&\equiv\!-\!\sum_{\sigma }
\big[{\cal C}^{( \sigma)}_{\alpha  }(\omega-{\cal L}_{\s})\hat O\big]\ti{d}^{\bar\sigma} ,
\\
\label{calwomega}
{\cal W}^{>}_{\alpha}(\omega)\hat O&\equiv\sum_{ \sigma }
\big[ \ti{d}^{\bar\sigma} ,c^{\sigma}_{\alpha }(\omega-{\cal L}_{\s})
   (\hat{d}^{\sigma}  \hat O) \big],
\\
{\cal W}^{<}_{\alpha}(\omega)\hat O&\equiv
 \sum_{ \sigma  }  \big[ \ti{d}^{\bar\sigma} ,c^{\bar\sigma\ast}_{\alpha  }({\cal L}_{\s}-\omega)
   (\hat O \hat  d^{\sigma} )\big],
\end{align}
\esube
and the propagator is given by
\be\label{Gomega}
 {\cal G}(\omega)=[i({\cal L}_{\s}-\omega)+\Sigma(\omega)]^{-1},
\ee
with
\be\label{sigmaw}
 \Sigma(\omega)\hat O=\sum_{\alpha\sigma }
\big[  \hat  d^{\bar\sigma} ,{\cal C}^{(\sigma)}_{\alpha  }(\omega-{\cal L}_{\s})
  \hat O  \big].
  \ee
Here [cf.\,\Eq{calCt}]
\begin{align}\label{calCw}
 {\cal C}^{(\sigma)}_{\alpha }(\omega)  \hat O
 \!=\!\! \sum 
 \! \big[c^{\sigma}_{\alpha }(\omega)(\hat{d}^{\sigma}\!\hat O)
 \!- c^{\bar\sigma\ast}_{\alpha }( -\omega)(\hat O \hat  d^{\sigma})\big],
 \end{align}
with  
$c^{\sigma}_{\alpha }( \omega)\equiv
\int_{0}^{\infty}\!dt\, 
e^{i\omega t}c^{\sigma}_{\alpha }(t)$.
It is  
obtained as \cite{Jin14244111,Shi16095002,Jin16083038,Xu22064130}
\begin{align}\label{appcw}
c^{(\pm)}_{\alpha }(\omega)
=\frac{1}{2}\big[
 \Gamma^{(\pm)}_{\alpha }(\mp \omega)
 +i\Lambda^{(\pm)}_{\alpha }(\mp \omega)\big],
\end{align}
where $\Gamma^{\pm}_\alpha(\omega)=f^{\pm}_\alpha(\omega)\Gamma_\alpha(\omega)$
with $f^{+}_\alpha(\omega)$ the Fermi-function of $\alpha$-lead and $f^{-}_\alpha(\omega)=1-f^{+}_\alpha(\omega)$.
In \Eq{appcw}, the real and the imaginary parts are the so-called bath
interaction spectrum and {\it dispersion}, respectively \cite{Xu029196}.
They are related via the Kramers-Kronig relations,
$\Lambda^{(\pm)}_{\alpha  }(\omega)
={\cal P}\int^\infty_{-\infty}\frac{{\mathrm d}\omega'}{2\pi}
\frac{1}{\omega\pm\omega'}\Gamma^{(\pm)}_{\alpha  }(\omega)$.
For the consideration of the Lorentzian-type form
of the hybridization spectral density, i.e.,
$
\Gamma_{\alpha}(\omega)=\frac{\Gamma_{\alpha } W^2_{\alpha}}{(\omega-\mu_\alpha)^2+W^2_{\alpha}},
$
with the coupling strength $\Gamma_{\alpha  }$ and the bandwidth $W_{\alpha}$
of lead-$\alpha$,
one can obtain
\begin{align}
\label{Lamb}
\Lambda^{(\pm)}_{\alpha  }(\omega)
 &=\frac{\Gamma_{\alpha  }(\omega)}{2\pi}
\Bigg\{{\rm Re}\left[\Psi\left(\frac{1}{2}
+i\frac{\beta(\omega-\mu_\alpha)}{2\pi}\right)\right]
\nla
-\Psi\left(\frac{1}{2}+\frac{\beta W_\alpha}{2\pi}\right)
\mp\pi\frac{\omega-\mu_\alpha}{W_\alpha}\Bigg\},
\end{align}
where ${\cal P}$ denotes the principle value of the integral,
and $\Psi(x)$ is the digamma function. 
 


\end{document}